\documentstyle[12pt,twoside,fleqn,espcrc1,epsf]{article}

\def\Journal#1#2#3#4{{#1} {\bf #2}, #3 (#4)}

\def\NP{{Nucl.~Phys.}}
\def\PL{{Phys.~Lett.}}

\def\PRD{{Phys.~Rev.~D}}
\def\PRL{Phys.~Rev.~Lett.}
\def\ZP{{Z.~Phys.}}

\newcommand{\bra}[1]{\langle #1 |}
\newcommand{\ket}[1]{|#1\rangle}

\newcommand{\T} {\mbox{T}}

\newcommand{\AmS}{{\protect\the\textfont2
  A\kern-.1667em\lower.5ex\hbox{M}\kern-.125emS}}

\title{Interactions of quarkonium at low energies}

\author{H. Fujii%
\address{Institute of Physics, University of Tokyo at Komaba,
              Tokyo 153-8902, Japan.}
        and 
        D. Kharzeev%
\address{RIKEN BNL Research Center, Brookhaven National Laboratory,\\
        Upton, NY 11973, USA}}

\begin{document}
\maketitle

\begin{abstract}
We present two examples of the short-distance QCD calculations for the
quarkonium ($\Phi$) interactions; 1)
$\pi \Phi$ elastic and $\pi \Phi \to \pi \Phi'$ cross sections,
and 2) the potential between two $\Phi$'s. 
The former is relevant for the $\Phi$ suppression in heavy ion
collisions; we find that the corresponding cross sections are very small. 
For the latter we derive the sum rule which relates the strength of 
the potential to the energy density of the QCD vacuum.
The key starting point in both cases is that the leading operator 
in the operator product expansion (OPE) for the low--energy scattering
amplitude is the trace of the
QCD stress tensor whose matrix elements are unambiguously 
fixed by the low energy theorem. 
The resonances in the two-pion scalar channel are taken into account
as a formfactor. 
\end{abstract}

\section{INTRODUCTION}

The heavy quarkonium (for which we will use a generic notation $\Phi$) 
has a small size 
($r\sim 1/(\alpha_s m_Q)$)
 and a large binding energy ($\epsilon \sim \alpha_s^2 m_Q$).
When we consider an interaction with light hadrons ($h, h'$) at a low
 (compared to $\epsilon$) energy, the amplitude 
can be expanded in multipoles \cite{pQCD,BP,Kai}: 
${\cal M} = \sum_i c_i \bra{h'}O_i(0)\ket{h}$,
where $c_i$ are the Wilson coefficients (polarizabilities) which
reflect the structure of the quarkonium. The matrix elements
of the gauge-invariant local operators $O_i(x)$ over the light hadron
state contain the information about the long distance part of the process;
we assume that the factorization scale in this formula is $\epsilon$.

This approach has been applied to the evaluation of inclusive cross 
sections of heavy quarkonium dissociation in hadron gas in \cite{KS}. 
Here we evaluate exclusive cross sections of $\pi \Phi$ interactions. 
Previously these interactions were addressed in 
Refs.~\cite{DK,SSZ,CS}; detailed description of our formalism and
results can be  found in Ref.~\cite{FK}.
  
The leading operator in OPE for the amplitude 
is $\frac{1}{2} g^2 {\bf E}^{a2}$, 
which describes the emission of two gluons in the color-singlet state.  
(The magnetic coupling is suppressed by the velocity $v\sim\alpha_s\ll
1$ for a heavy $\Phi$.)  
One can re-write this operator in the Lorentz covariant form,
$\frac{1}{2}g^2{\bf E}^{a2}
=\frac{4\pi^2}{9}\theta_\mu^\mu
+\frac{1}{2}g^2\theta^{(G)}_{00}$. As will be discussed below, the 
scale anomaly implies that  
the matrix elements of the stress tensor, 
$\theta_\mu^\mu=-(9g^2/32\pi^2)(G_{\mu\nu}^a)^2$ do not depend on the coupling 
$g^2$ and dominate over the matrix elements of the gluon tensor operator,
$\theta^{(G)}_{00}$, which is manifestly suppressed by $g^2$. In other words, 
scale anomaly effectively eliminates the factor of $g^2$ in the amplitude!

\section{$\pi \Phi$ INTERACTIONS}

\begin{figure}
\begin{minipage}{0.47\textwidth}
\epsfxsize=8cm
\epsffile{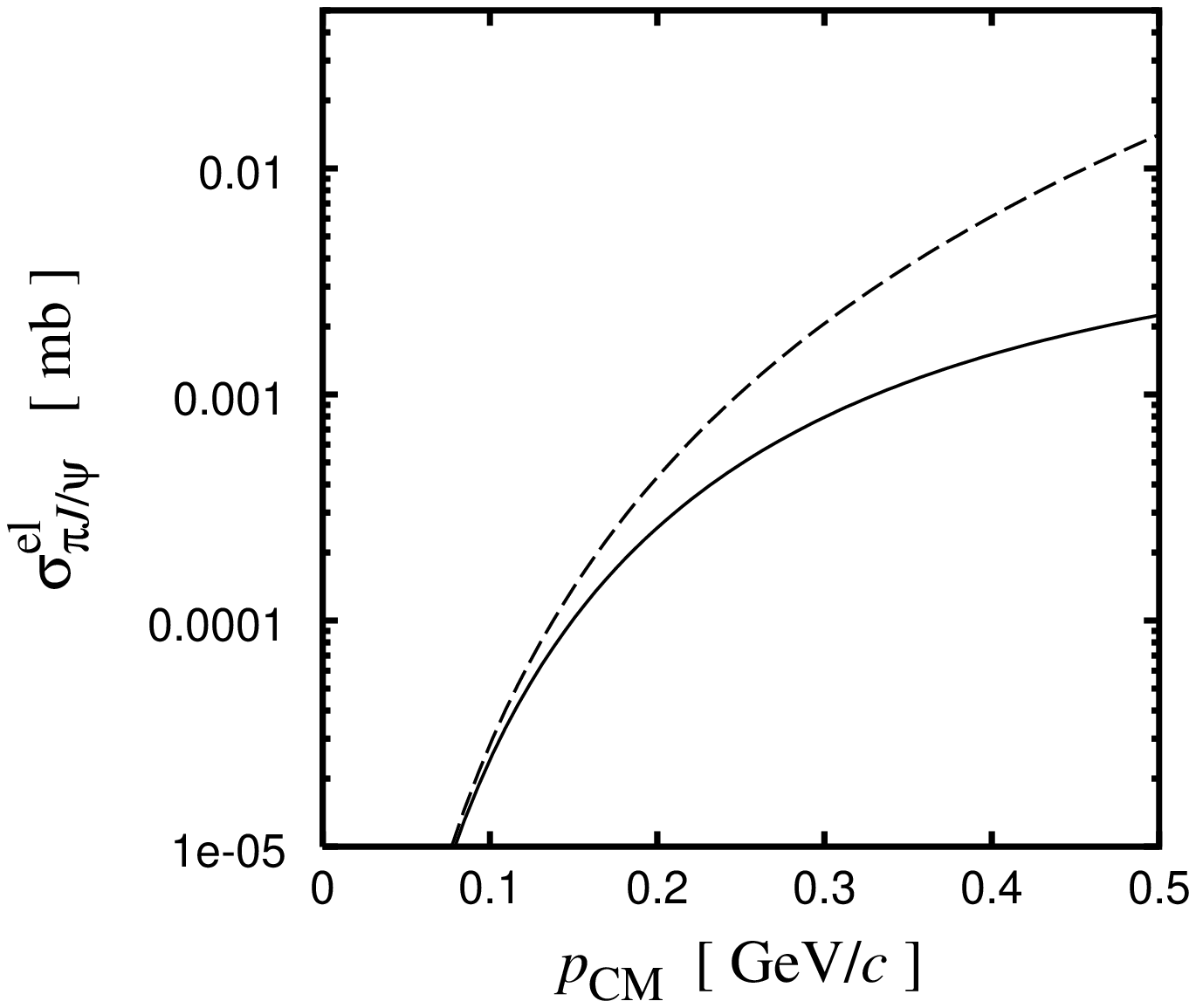}
\caption{$\pi J/\psi$ elastic cross section as a function of the CM momentum;
solid line is the complete result with the $\pi\pi$ formfactor $F$;
dashed line shows the case without $F$.}
\label{fig1}
\end{minipage}
\hfill 
\begin{minipage}{0.47\textwidth}
\epsfxsize=7.8cm
\epsffile{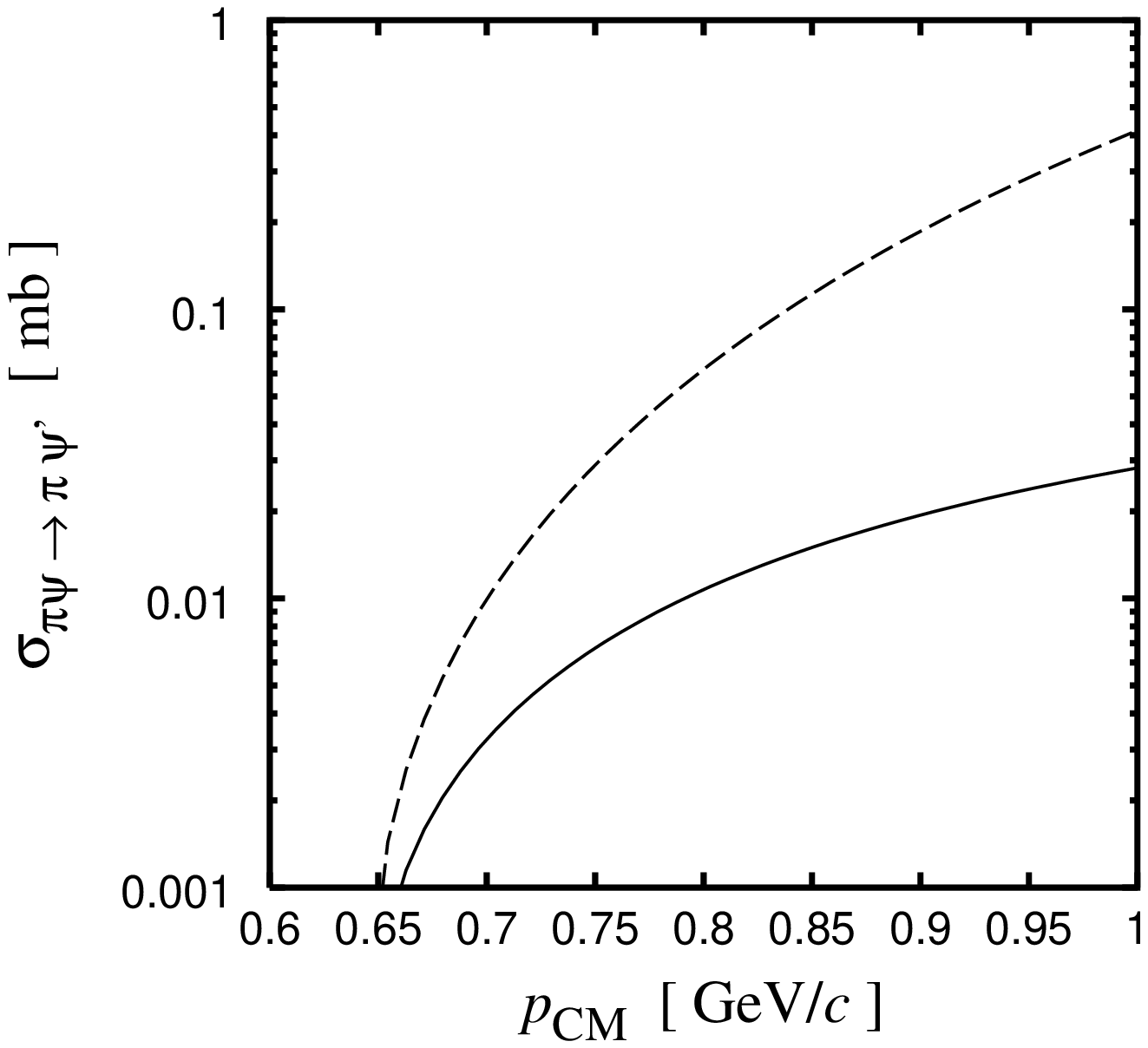}
\caption{$\pi J/\psi \to \pi \psi'$ transition cross section; for notations,
see the caption to Fig.~1.}
\label{fig2}
\end{minipage}
\end{figure}

Let us consider $\Phi$ interactions with pions, which are the
dominant degrees of freedom at low energies/temperatures.
Since $\theta_\mu^\mu$ is scale invariant, we can match it
onto its expression in terms of the effective pion
lagrangian, and find a beautiful relation \cite{matching} 
$\bra{\pi^+(p')}\theta_\mu^\mu \ket{\pi^+(p)}=(p'-p)^2=t$ valid in the
chiral limit.
In the leading order in OPE, 
the amplitude of the $\pi \Phi$ elastic 
scattering can be written as
\begin{equation}
{\cal M}^{\Phi \pi}= \bar d_2 \frac{a_0^2}{\epsilon_0} \frac{4\pi^2}{9}
\bra{\pi^+(p')} \theta_\mu^\mu \ket{\pi^+(p)}
=\bar d_2 \frac{a_0^2}{\epsilon_0} \frac{4\pi^2}{9}
t F(t),
\end{equation}
where, in the polarizability $\bar d_2 a_0^2 / \epsilon_0 $, we
have factored out the Bohr radius $a_0$ and the Rydberg energy
$\epsilon_0$ of quarkonium, respectively; 
$F(t)$ is 
the pion formfactor in the scalar-isoscalar channel, which takes 
account of the resonances in this channel.

The resulting elastic cross section of $\Phi$ with mass $M$
\begin{equation}
\sigma^{\rm el}(s)=\frac{1}{4\pi s} \frac{M^2}{4 p_{\rm cm}^2}
\left ( \bar d_2 \frac{a_0^2}{\epsilon_0} \right )^2
\left ( \frac{4\pi^2}{9} \right )^2 
\int_0^{4p_{\rm cm}^2} d(-t) t^2 |F(t)|^2
\end{equation}
appears to be very small (see Fig.~1) because of the factor
$a_0^{4}/\epsilon_0^2$ stemming from the double-dipole nature of the
interaction. Moreover, it is further suppressed by the dependence on the 
momentum transfer $t$ of the cross section dictated by 
the chiral symmetry.

In Fig.~\ref{fig1} we show the $\pi \Phi$ elastic cross section 
for the $J/\psi$ case; one can see that the cross section is very 
small. 
This result implies that the $J/\psi$'s produced in heavy ion collisions 
interact with the surrounding $\pi$ gas very weakly. 
The $p_T$ broadening of the $\Phi$'s by the interactions inside the 
pion gas therefore is negligible.

Next we consider the inelastic excitation process, 
$\pi \Phi \to \pi \Phi'$. The
energy transfer in this process is on the order of the binding 
energy, $\Delta=M'-M=O(\epsilon_0)$, which may invalidate our starting
assumption of the factorization. Fortunately the double-dipole form
is intact, but {\it non-local} in time.
We replace the gluon energy appearing in the
energy denominator with the typical value, $\Delta$, as an 
approximation, to make the gluonic operator local:
\begin{equation}
{\cal M}^{\pi \Phi\to \pi \Phi'} 
\simeq 
\frac{1}{3N_c} \bra{\phi '}r^i \frac{1}{H_a+\epsilon-\Delta}r^i \ket{\phi}
\bra{\pi}\frac{1}{2} g^2 {\bf E}^2 \ket{\pi}
\equiv \bar d'_2 \frac{a_0^2}{\epsilon_0} 
\bra{\pi}\frac{1}{2} g^2 {\bf E}^2 \ket{\pi},
\end{equation}
where the $\phi$ ($\phi'$) is the internal wave function of $\Phi$ 
($\Phi'$), and $H_a(r)$ is the effective hamiltonian describing the
intermediate, $SU(N_c)$ color--adjoint quark--anti-quark state.
With this amplitude we get the transition
cross section shown in Fig.~2.

Our replacement of the gluon energy in the energy denominator is an 
approximation with an accuracy that cannot be evaluated 
{\it a priori}. We can, however, apply our formula to the decay process
$\psi' \to J/\psi \pi \pi$
and compare our results, $\Gamma^{\pi\pi}$=260 (70) keV obtained with
(without) taking account of the formfactor $F(s)$,
 with the experimentally measured value of 135$\pm$20 keV. 
From this comparison 
we conclude that our calculations are probably valid  up to a factor of 2. 

Our results support the idea \cite{MS1,KS} that the interactions 
of heavy quarkonia in the hadron gas and in the quark-gluon plasma 
are very different.

\section{ONIUM-ONIUM SCATTERING AT LOW ENERGY}

Low-energy onium-onium scattering is a very interesting subject, 
which was addressed by several authors
\cite{Ap,BP,BM}. 
One may regard this problem as a first step toward the understanding 
of nuclear force from QCD.
Here we derive a sum rule for this interaction \cite{FK}.

After the multipole expansion for the interactions of both $\Phi$'s with 
the gluon field is performed, the potential
between them can be expressed through the $\theta_\mu^\mu$ correlator:
\begin{eqnarray}
V(R)=-{\cal M}^{\Phi\Phi}(R)&=& -i\int_\infty^\infty dt 
\left ( \bar d_2 \frac{a_0^2}{\epsilon_0}\right )^2
\left (\frac{4 \pi^2}{b} \right )^2
\bra{0} \T \theta_\mu^\mu(x) \theta_\nu^\nu(0)\ket{0}
\nonumber \\
&=&-\left ( \bar d_2 \frac{a_0^2}{\epsilon_0}\right )^2
\left (\frac{4\pi^2}{b}\right )^2
\int_0^\infty d\sigma^2 \rho_\theta(\sigma^2)
\frac{1}{4\pi R}e^{-\sigma R},
\end{eqnarray}
where we have introduced the spectral function for the
$\theta_\mu^\mu$ correlator as
$\bra{0}\T\theta_\mu^\mu(x)\theta_\nu^\nu(0)\ket{0}=
\int d\sigma^2 \rho_\theta(\sigma^2)\Delta_F(x;\sigma^2)$ with the
coordinate-space scalar Feynman propagator, $\Delta_F$;
$b=(11N_c-2N_f)/3=9$. 

An important theorem \cite{nsvz} 
for the $\theta_\mu^\mu$ correlator relates it to the 
non-perturbative energy density of the QCD vacuum: 
\begin{equation}
i \int d^4 x e^{iqx}
\bra{0}\T\theta_\mu^\mu(x)\theta_\nu^\nu(0)\ket{0}=
\int d\sigma^2 \frac{\rho_\theta(\sigma^2)}{\sigma^2-q^2-i\epsilon}
\stackrel{q\to 0}{\longrightarrow} -4 \bra{0}\theta_\mu^\mu(0)\ket{0}
=-16 \epsilon_{\rm vac} .
\end{equation}
Since the vacuum energy density is divergent in perturbation
theory, we define the r.h.s. by subtracting the perturbative part. This
relates the integral of the spectral density to $\epsilon_{\rm vac}$
as $\int (d\sigma^2 /\sigma^2) 
(\rho_\theta^{\rm phys}(\sigma^2)-\rho_\theta^{\rm pt}(\sigma^2))
=-16 \epsilon_{\rm vac}$.
In the heavy quark limit, this relation leads to the following 
sum rule for the potential 
\begin{equation}
\int_r^\infty d^3 {\bf R} (V_\theta(R)-V_\theta^{\rm pt}(R))=
-\left (\bar d_2 \frac{a_0^2}{\epsilon_0}\right )^2
\left (\frac{4 \pi^2}{b} \right )^2
16 |\epsilon_{\rm vac}|;
\end{equation}
where $r$ is the size of the onium. This sum rule relates the 
overall strength of the interaction between the dipoles to the 
energy density of the QCD vacuum. 

The short-- and 
long--distance limits of the potential are analyzed in Ref.~\cite{FK}. 
In particular, it appears that the long-distance limit of 
the potential has an 
interesting $R$ dependence, $\sim R^{-5/2} exp(-2 \mu_{\pi} R)$ 
($\mu_{\pi}$ is the pion mass), 
with a non-trivial dependence on the numbers of colors $N_c$ and 
flavors $N_f$, $\sim (N_f^2-1)^2/(11 N_c - 2 N_f)^2$.

\end{document}